\documentclass[12pt]{article}
\usepackage{graphicx}

\newcommand\pubnumber{NuPhys2016-Parsa}
\newcommand\pubdate{\today}

\def\support{\footnote{This project has received funding from the European Union's Horizon 2020 Research and Innovation programme under grant agreement No. 654168.}}

\renewcommand{\thefootnote}{$\star$}

\def\Title#1{\begin{center} {\Large #1 } \end{center}}
\def\Author#1{\begin{center}{ \sc #1} \end{center}}
\def\Address#1{\begin{center}{ \it #1} \end{center}}

\newcommand\pubblock{\rightline{\begin{tabular}{l} \pubnumber\\
         \pubdate  \end{tabular}}}
\newenvironment{Abstract}{\begin{quotation}  }{\end{quotation}}
\newenvironment{Presented}{\begin{quotation} \begin{center} 
             PRESENTED AT\end{center}\bigskip 
      \begin{center}\begin{large}}{\end{large}\end{center} \end{quotation}}

\def\beq{\begin{equation}}
\def\eeq#1{\label{#1}\end{equation}}
\def\eeqn{\end{equation}}

\def\beqa{\begin{eqnarray}}
\def\eeqa#1{\label{#1}\end{eqnarray}}
\def\eeqan{\end{eqnarray}}

\let\bar=\overbar

\def\Dslash{\not{\hbox{\kern-4pt $D$}}}
\def\dslash{\not{\hbox{\kern-2pt $\del$}}}

\def\msb{{\bar{\ssstyle M \kern -1pt S}}}

\begin{document}
\begin{titlepage}
\pubblock

\vfill
\Title{Baby MIND Experiment Construction Status}
\vfill

\def\Author#1 {\begin{center}{ \sc #1} \end{center}}
\def\Address#1 {\begin{center}{ \it #1} \end{center}}

\small{
\Author{M.~Antonova$^{1}$, R.~Asfandiyarov$^{2}$, R.~Bayes$^3$, P.~Benoit$^4$,  A.~Blondel$^2$, M.~Bogomilov$^5$, A.~Bross$^6$, F.~Cadoux$^2$, A.~Cervera$^{7}$, N.~Chikuma$^{8}$, A.~Dudarev$^4$, T.~Ekel\"of$^9$, Y.~Favre$^2$, S.~Fedotov$^1$,  S-P.~Hallsj\"o$^{3}$, A.~Izmaylov$^1$ , Y.~Karadzhov$^2$, M.~Khabibullin$^{1}$, A.~Khotyantsev$^1$, A.~Kleymenova$^{1}$, T.~Koga$^{8}$, A.~Kostin$^{1}$, Y.~Kudenko$^{1}$,   V.~Likhacheva$^{1}$, B.~Martinez$^2$, R.~Matev$^5$, M.~Medvedeva$^{1}$, A.~Mefodiev$^{1}$, A.~Minamino$^{10}$, O.~Mineev$^1$, M.~Nessi$^4$, L.~Nicola$^2$, E.~Noah$^2$,  T.~Ovsiannikova$^{1}$, H.~Pais Da Silva$^4$, S.~Parsa$^2$\support , M.~Rayner$^4$, G.~Rolando$^4$, A.~Shaykhiev$^1$, P.~Simion$^9$ , F.J.P.~Soler$^3$,  S.~Suvorov$^{1}$, R.~Tsenov$^5$, H.~Ten Kate$^4$, G.~Vankova-Kirilova$^5$ and N.~Yershov$^1$.}

\Address{$^1$Institute of Nuclear Research, Russian Academy of Sciences, Moscow, Russia.\\
$^2$University of Geneva, Section de Physique, DPNC, Geneva, Switzerland.\\
$^3$University of Glasgow, School of Physics and Astronomy, Glasgow, UK.\\
$^4$European Organization for Nuclear Research, CERN, Geneva, Switzerland.\\
$^5$University of Sofia, Department of Physics, Sofia, Bulgaria.\\
$^6$Fermi National Accelerator Laboratory, Batavia, Illinois, USA.\\
$^{7}$IFIC (CSIC $\&$ University of Valencia), Valencia, Spain.\\
$^8$University of Tokyo, Tokyo, Japan.\\
$^9$University of Uppsala, Uppsala, Sweden.\\
$^{10}$Yokohama National University, Yokohama, Japan.
}
}

\vfill
\def\Author#1{\begin{center}{ \sc #1} \end{center}}
\def\Address#1{\begin{center}{ \it #1} \end{center}}

\begin{Abstract}
\small{
Baby MIND is a magnetized iron neutrino detector, with novel design features, and is planned to serve as a downstream magnetized muon spectrometer for the WAGASCI experiment on the T2K neutrino beam line in Japan. One of the main goals of this experiment is to reduce systematic uncertainties relevant to CP-violation searches, by measuring the neutrino contamination in the anti-neutrino beam mode of T2K. Baby MIND is currently being constructed at CERN, and is planned to be operational in Japan in October 2017.
}
\end{Abstract}

\vfill
%\small{
\begin{Presented}
NuPhys2016, Prospects in Neutrino Physics\\
Barbican Centre, London, UK,  December 12--14, 2016
\end{Presented}
%}
\vfill
\end{titlepage}
\def\thefootnote{\fnsymbol{footnote}}
\setcounter{footnote}{0}

\section{Introduction}

The Baby MIND experiment as the downstream MRD of the J-PARC ~T59 experiment (WAGASCI) \cite{wagasci}, was approved by the CERN Research Board in December 2015 \cite{spsc}. The WAGASCI experiment, with a novel detector arragement which creates arrays of small cubic cells whose walls are made from plastic scintillators, provides better acceptance compared to a classical layout of successive X and Y planes scintillator bars. The adoption of WAGASCI units for upgrades to the T2K ND280 near detector is being considered, motivated by the need to reduce systematics due to nuclear effects in water, one of the dominant systematic uncertainties in the T2K neutrino oscillation analyses, and of direct relevance to the HyperK project \cite{hyperK}.

\begin{figure}[b]
\centering
\includegraphics[height=4cm]{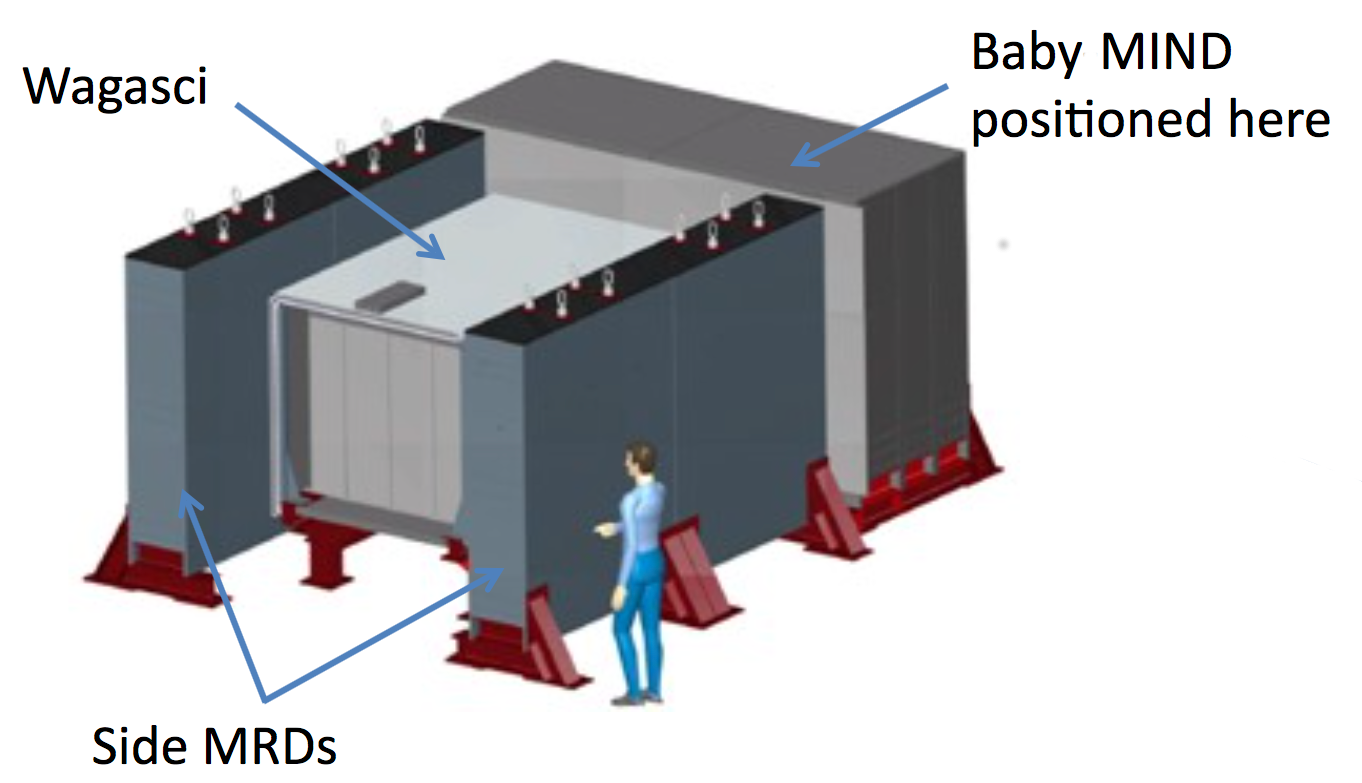}
\includegraphics[height=4cm]{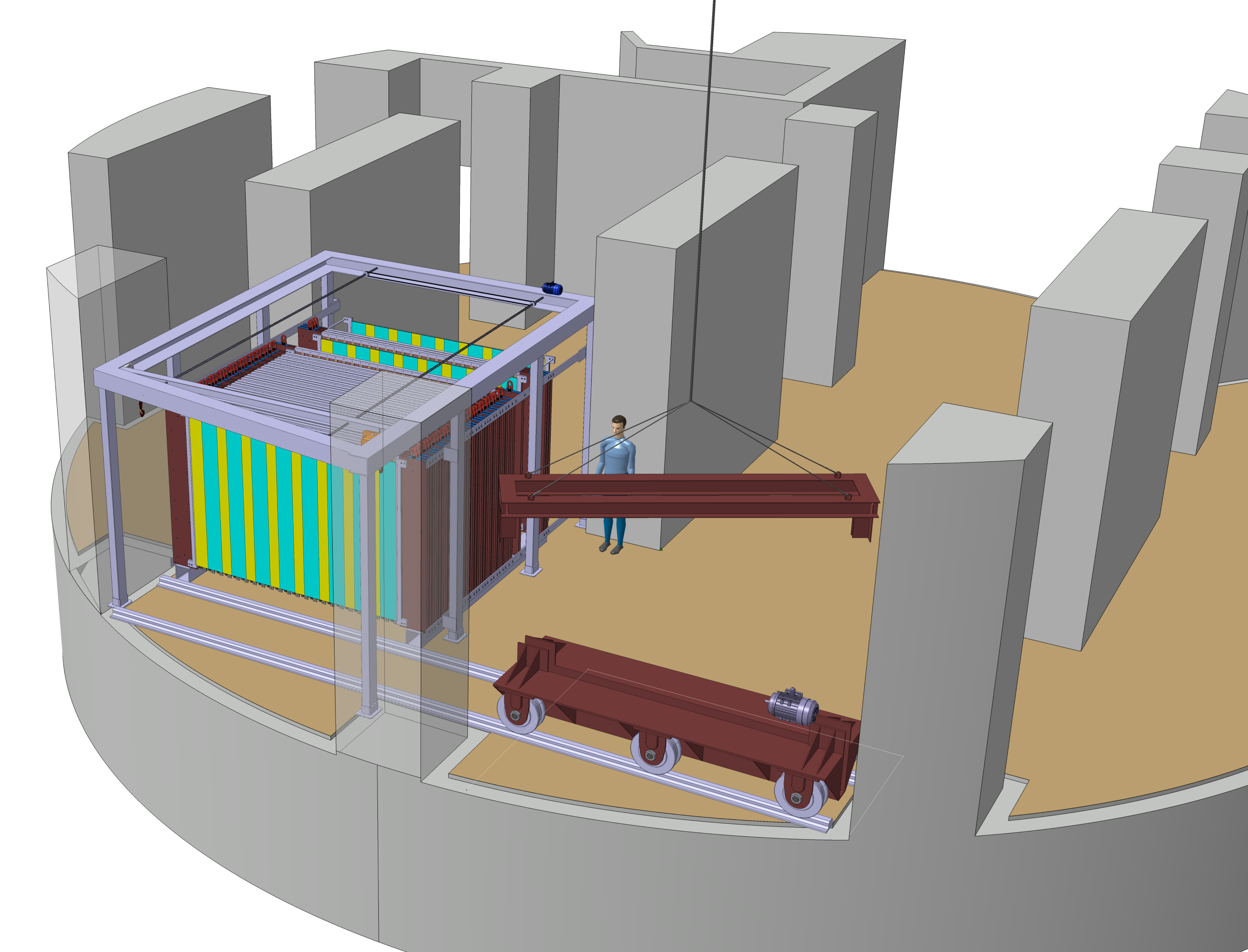}
\caption{\label{fig1} \small{(Left) Sketch of WAGASCI detector, side MRDs and Baby MIND, (Right) Baby MIND installation in ND280 pit.}}
\end{figure}

One challenge to be addressed by the Baby MIND collaboration is that of obtaining high charge identification efficiencies for $\mu^+$ and $ \mu^-$ down to 400 MeV/c. Magnetized iron neutrino detectors are limited by multiple scattering in the iron, and their use is overlooked for applications requiring good charge ID efficiencies below 1 GeV/c. By optimising the distance between the first magnet modules, rendered possible by the magnet design, our simulations show high charge identification efficiencies of $97\%$ down to 500 MeV/c, and capabilities for charge identification down to 300 MeV/c.

\section{Baby MIND Detector design}
\begin{figure}
\centering
\includegraphics[height=4cm]{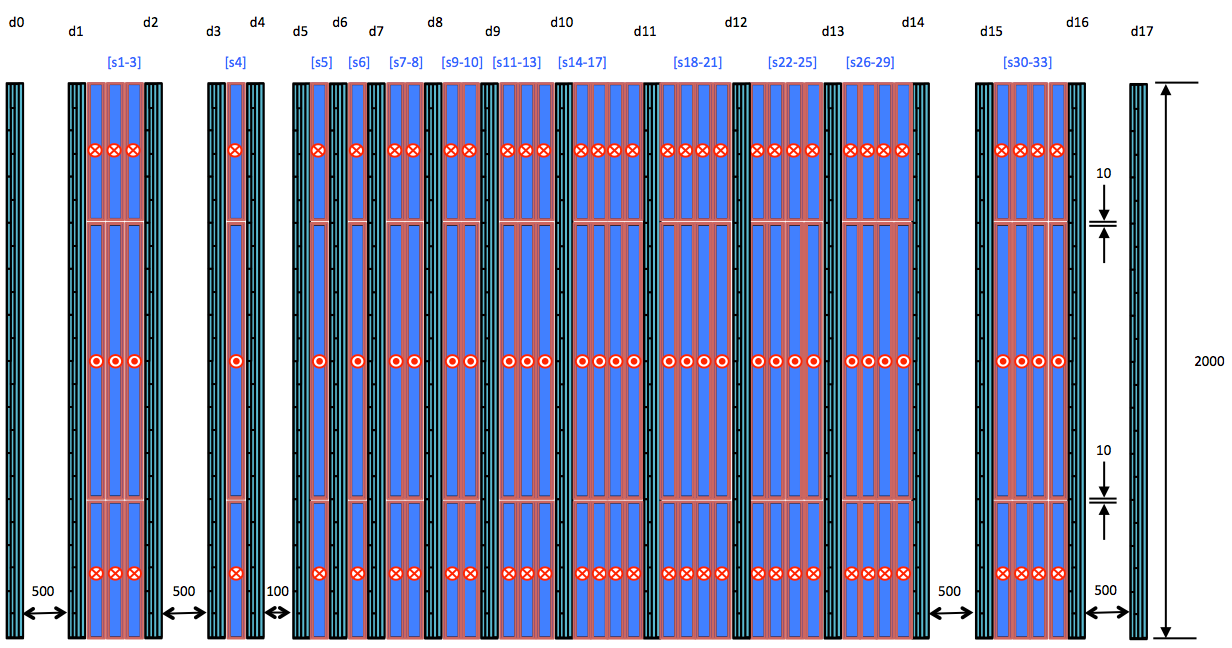}
\caption{\label{fig2layout} \small{One possible Baby MIND layout, with muons incident from the left. Scintillator modules are referenced d0 to d17, magnet modules are referenced s1 to s33.}}
\end{figure}

The design and construction of the Baby MIND detector was very much constrained by the need to operate the detector both at CERN and in Japan on a relatively short timescale, and also the installation limitations in Japan via a narrow shaft (Figure\ref{fig1}), which in particular has driven the new magnetization scheme of the detector. 
Baby MIND is built from sheets of iron interleaved with detector modules as shown in Figure \ref{fig2layout}, but unlike traditional layouts for magnetized iron neutrino detectors (e.g.MINOS) which tend to be monolithic blocks with a unique pitch between consecutive steel segments and large conductor coils threaded around the whole magnet volume, the Baby MIND iron segments are all individually magnetized, allowing for far greater flexibility in the setting of the pitch between segments, and in the allowable geometries that these detectors can take. This is of relevance for neutrino experiments that might consider muon spectrometers based on magnetized iron surrounding an active detector volume such as plastic scintillator, water cherenkov, liquid argon.

\section{Detector hardware construction}

The different detector hardware components and systems are listed below:
\small{
\begin{itemize}
\item  Magnet module.
\item  Scintillator detector module.
\item  Cable bundles.
\item  Front-end electronics module. 
\item  DAQ system.
\item  Mechanics support systems.
\end{itemize}
}
\normalsize
The general approach of the collaboration is to ensure documentation and traceability of components and systems. A construction database has been written to store parameters relevant to the detector hardware such as detector geometrical configurations, serial numbers of components and test data. The majority of systems have undergone prototype validation before launching the production phases. Assembly and qualification procedures are drafted for each system. Integration procedures ensure the correct handling of interfaces during systems integration phases.
\small
\subsection{Magnet modules status}
\begin{figure}[t]
\centering
\includegraphics[height=4cm]{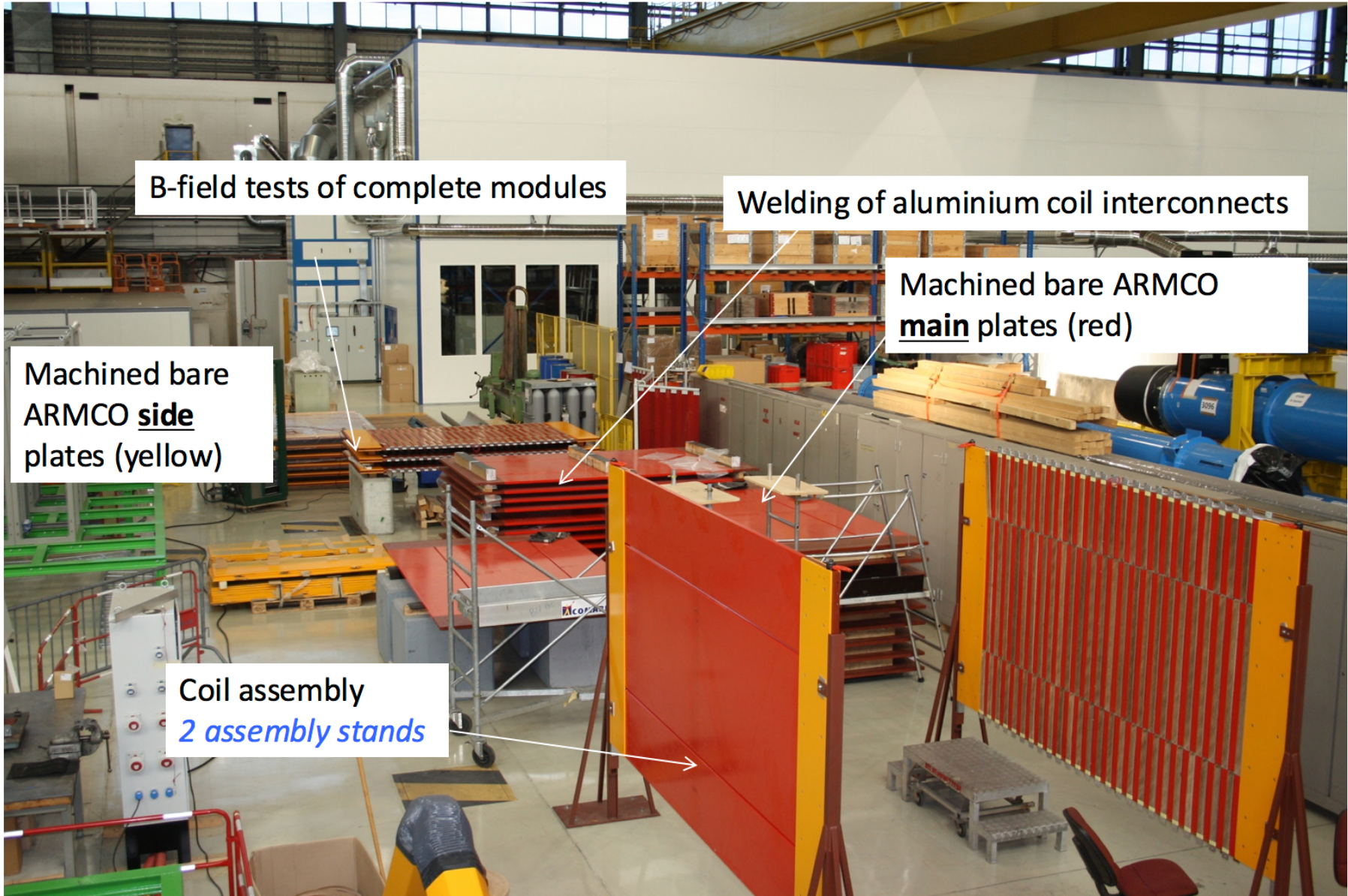}
\includegraphics[height=4cm]{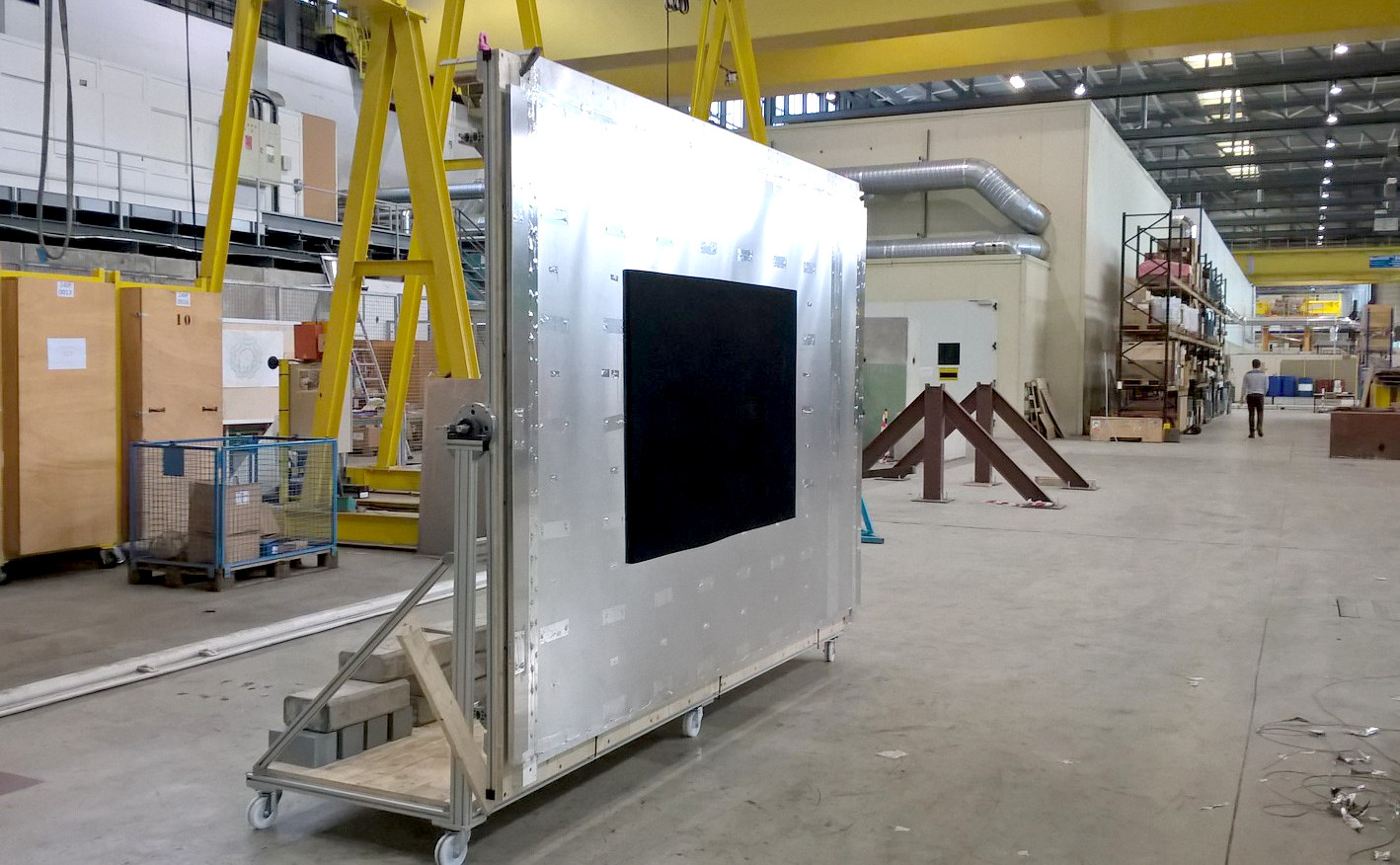}
\caption{\label{fig2} \small{(Left) The Magnet assembly zone, (Right) A scintillator module.}}
\end{figure}

The magnetization scheme for Baby MIND was designed at CERN\cite{magnet}. Construction of the magnet modules was completed there by February 2017. The magnet modules are made from ARMCO steel sheets with two horizontal slits machined in the center and are wrapped by aluminium coil in a sewing pattern, Figure \ref{fig2}.
This design allows the magnetic flux to be very uniform in the central tracking region, and contains the return flux in the side planes. The required horizontal magnetic field of $1.5$ T can be reached for a current of $140$ A and a dissipation power of $350$ W per module.

\subsection{Scintillator modules status}
The scintillator modules with dimensions of $2 \times 3$  m$^2$ contain 95 horizontal bars and 16 vertical bars each. The bars are extruded scintillators with wavelength shifting fiber and custom photosensor connectors manufactured under the responsibility of INR \cite{INR}. One completed scintillator module is shown in Figure \ref{fig2}.
A total of 18 standard detector modules will be used in Baby MIND final layout, interleaved with magnet modules. The final position of these modules with respect to magnet modules is not frozen yet and is subject to changes depending on different performance priorities. 
The production of all the units is scheduled to be finished for the end of April 2017.
  
  \subsection{Electronics and DAQ system}
\begin{figure}
\centering
\includegraphics[height=6.2cm]{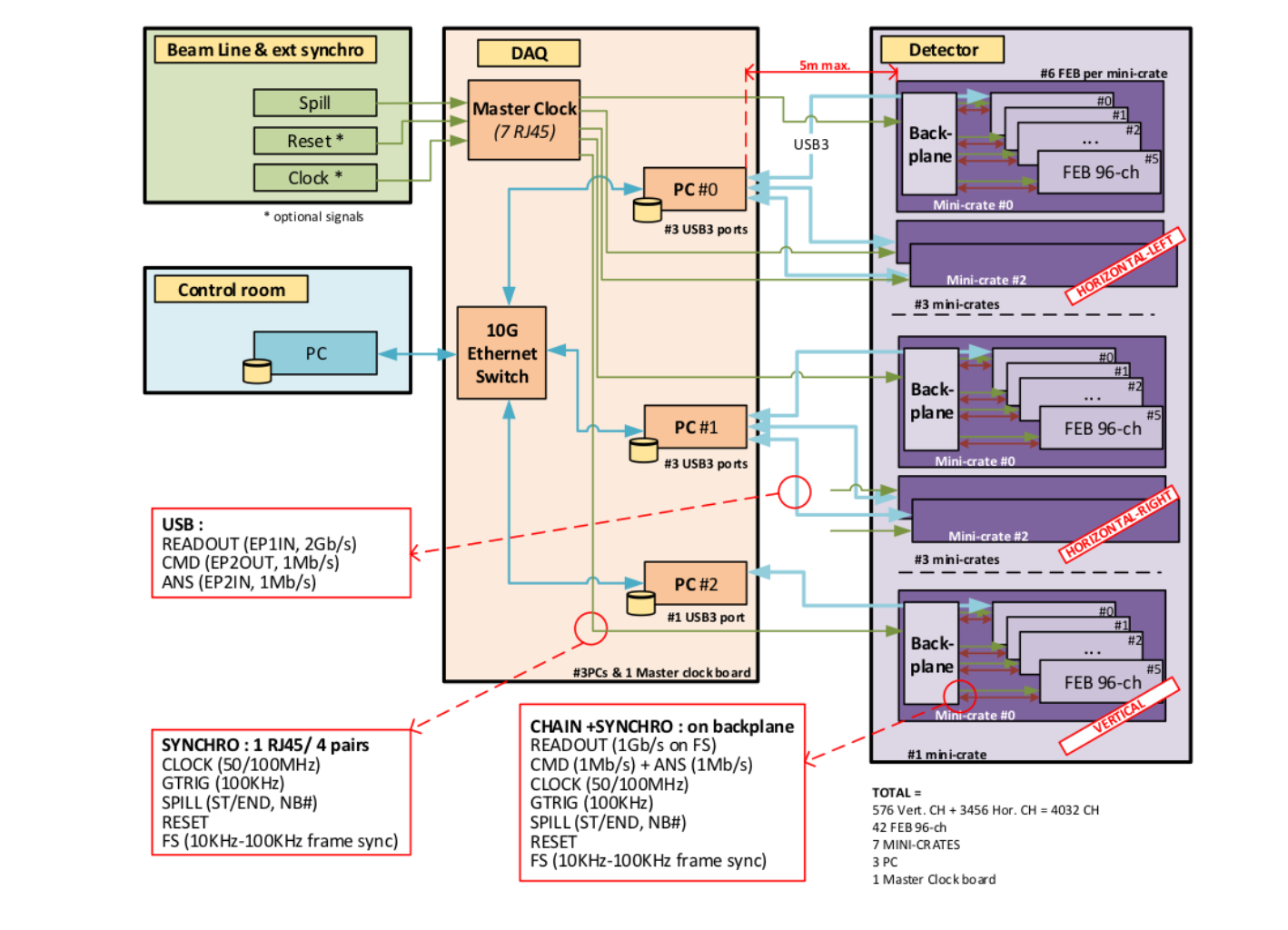}
\includegraphics[height=6.2cm]{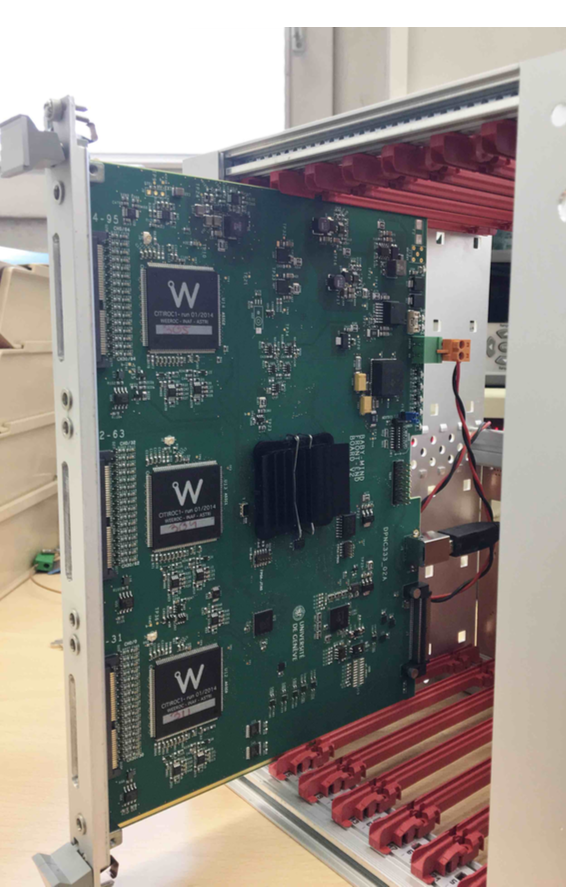}
\caption{\label{fig3} \small{(Left) The electronic readout scheme, (Right) FEBv2.}}
\end{figure}

\small
The Baby MIND electronic readout scheme has been published in ref \cite{electronics}. The Front End Board for the experiment was developed by the University of Geneva based on the CITIROC ASIC. The readout system includes two additional ancillary boards, the Backplane and the Master Clock Board (MCB). The readout system architecture is shown in Figure \ref{fig3}.  After the extensive validation tests of the FEBv1 during beam tests at the T9 beamline at CERN with $1$ to $10$ GeV/c muons in summer 2016, FEBv2 was developed, integrating layout changes dictated by cabling considerations and with additional features for daisy chaining and synchronisation of multiple boards.
The DAQ PCs are connected by USB3 links to the first FEB in a group of 6 FEBs that are daisy chained using RJ45 links.  
   
\normalsize
\section{Outlook}
The collaboration is currently focused on completion of the detector construction and preparation for the upcoming beam tests at the T9 beamline at CERN in May and June 2017. After full characterization, the detector will be shipped to Japan in July 2017. The installation and commissioning of the detector in J-PARC will take place in September 2017 in order to be ready for beam at J-PARC in October 2017.

%\fi

\bibliographystyle{unsrt}

\bibliography{WAGASCI-bibliography}
 
\end{document}